\documentclass[a4paper,11pt]{article}
\pdfoutput=1 
\usepackage{jheppub} 
\usepackage[T1]{fontenc} 
\usepackage{amsmath,amssymb,amsfonts}
\usepackage{graphicx}
\usepackage{enumitem}
\usepackage[usenames,dvipsnames]{color}
\usepackage{verbatim}
\usepackage[percent]{overpic}
\usepackage{rotating}
\usepackage{hyperref}
\usepackage{xcolor}
\usepackage{array}
\usepackage{mathtools}
\usepackage{lmodern}
\usepackage[normalem]{ulem}
\usepackage{braket}
\usepackage{tensor}
\usepackage{dsfont}
\usepackage{bm}
\usepackage{tikz}
\usetikzlibrary{decorations.pathmorphing}

\usepackage{CJKutf8}

\newcommand{\tcb}[1]{\leavevmode{\color{Blue}{#1}}}

\newcommand{\dd}{\text{d}}

\newcommand{\rr}[1]{\left(#1\right)}
\renewcommand{\H}{\hat{H}}
\newcommand{\openone}{\mathds{1}}
\newcommand{\sx}{\mathsf{x}}

\newcommand{\ii}{\mathsf{i}}
\newcommand{\pd}{\partial}

\DeclareMathOperator{\tr}{Tr}
\renewcommand{\bar}{\overline}
\title{Unruh-DeWitt detector in dimensionally-reduced static spherically symmetric spacetimes}

\author[a,b,1]{Erickson Tjoa,\note{Corresponding author.}}
\author[a,c]{Robert B. Mann}

\affiliation[a]{Department of Physics and Astronomy,
\\University of Waterloo, Waterloo, Ontario, N2L 3G1, Canada}
\affiliation[b]{Institute for Quantum Computing, \\University of Waterloo, Waterloo, Ontario, N2L 3G1, Canada}
\affiliation[c]{Perimeter Institute for Theoretical Physics,
\\Waterloo, Ontario, N2L 2Y5, Canada}

\emailAdd{e2tjoa@uwaterloo.ca}
\emailAdd{rbmann@uwaterloo.ca}

\abstract{We study the dynamics of an Unruh-DeWitt detector interacting with a massless scalar field in an arbitrary static spherically symmetric spacetimes whose metric is characterised by a single metric function $f(r)$. In order to obtain clean physical insights, we employ the derivative coupling variant of the Unruh-DeWitt model in (1+1) dimensions where powerful conformal techniques enable closed-form expressions for the vacuum two-point functions. Due to the generality of the formalism, we will be able to study a very general class of static spherically symmetric (SSS) background. We pick three examples to illustrate our method: (1) non-singular Hayward black holes, (2) the recently discovered $D\to 4$ limit of Gauss-Bonnet black holes, and (3) the ``black bounce'' metric that interpolates Schwarzschild black holes and traversable wormholes.  We also show that the derivative coupling Wightman function associated with the generalized Hartle-Hawking vacuum satisfies the KMS property with the well-known temperature $f'(r_\textsc{h})/(4\pi)$, where $r_\textsc{h}$ is the horizon radius.}

\begin{document} 
\maketitle
\flushbottom

\section{Introduction}

The utility  of particle detector models for probing fundamental physics within the framework of quantum field theory (QFT) in curved spacetimes is both well-known and well-established. First developed by Unruh and DeWitt \cite{Unruh1979evaporation,DeWitt1979}, they are useful because they represent localized observers interacting with quantum fields using non-relativistic two-level systems (qubits) in a manner analogous to how light-matter interactions in quantum optics are described. They are now refined to a point where finite-size effects are included in  a covariant manner \cite{Tales2020GRQO,Bruno2020time-ordering,Maria2021causality} and capture the impact of quantized centre-of-mass degrees of freedom of the detector \cite{Lopp2021deloc}. These models also admit ``variants'': these include non-linear interactions \cite{Tjoa2020resonance}, different spins of both the detector and the field \cite{Hummer2016bosonfermionZM}, harmonic oscillator-based detectors  \cite{Lin2007backreaction,Bruschi:2012rx,Brown2013harmonic,hotta2020duality}, spacetime superpositions \cite{Foo:2020jmi,Foo:2021gkl,Foo:2021fno}, quantum causal switches \cite{Henderson:2020zax}, 
and even experimental models using lasers \cite{Gooding2020interferometric} and non-linear optics \cite{Adjei2020nonlinear}. Within the subject matter now known as relativistic quantum information (RQI), particle detector models are now developed enough to provide a local measurement theory for QFT that respects causality \cite{polo2021detectorbased}. 

Two of the most well-known effects studied using detector models are  the Unruh and Hawking effects \cite{Unruh1979evaporation,hawking1975particle}. These effects emphasize the notion that fields are more fundamental than particles, since the ``particle content'' of the field is observer-dependent: for the Unruh effect, uniformly accelerating observers with proper acceleration $a$ interacting with the Minkowski vacuum experience\footnote{Of course, the word ``experience'' needs to be qualified, since local observers measure local temperature that may depend on their states of motion.} a thermal bath at temperature $\mathsf T_\textsc{u}\propto a$; for the Hawking effect, stationary observers interacting with the Hartle-Hawking vacuum of a Schwarzschild black hole of mass $M$ experience a thermal bath with temperature $\mathsf  T_\textsc{h}\propto M^{-1}$. In both cases, particle detectors can be used to ``certify'' that this is the case in an operational manner. We note in passing that while the Unruh and Hawking effects are ``stationary'' effects, requiring certain kinds of time-translation symmetry in the scenario,  particle detector models are versatile enough to study time-dependent situations, such as when the background spacetime is an expanding universe \cite{Gibbons1977cosmological,Steeg2009,Simidzihja2017cosmo,Blasco2015Huygens,Blasco2016broadcast,bibhas2020}. 

Somewhat unfortunately, the versatility of these particle detector models (hereafter collectively called \textit{Unruh-DeWitt \emph{(UDW)} detector models}) leads to one big shortcoming when it comes to practical calculations: with the exceptions of highly symmetric situations such as (conformally) flat spacetimes, the detector-field interaction is often not explicitly solvable even within first-order perturbation theory. This is already the case even for Schwarzschild geometry: the ``greybody factor'' that originates from the effective potential in the radial direction renders the calculations difficult even numerically  \cite{Hodgkinson4Dschwarzschild,Ng2014Schwarzschild,Casals2020communication}, although in \cite{Hodgkinson4Dschwarzschild} it was shown that one can nonetheless obtain the so-called \textit{detailed balance condition} for the excitation-to-deexcitation (EDR) ratio of the two-level detector. For this reason, UDW detectors in (1+1)-dimensional ``dimensionally-reduced'' models have been extensively used in the literature (see, e.g., \cite{birrell1984quantum} and references therein). Using (1+1)-dimensional truncation, the Wightman two-point functions for massless fields can be solved \textit{exactly} in closed form due to the conformal flatness of all two-dimensional geometries, and the massive field counterpart does not pose too much trouble.

Of course, (1+1)-dimensional models extract another price for their exact solvability: in many cases, such as (truncated) Schwarzschild geometry, it is well-known that massless fields exhibit infrared (IR) divergences and the (vacuum) Wightman two-point functions do not exhibit the short-distance behaviour expected of the physically reasonable class of states in (3+1)-dimensional spacetimes called \textit{Hadamard states}. The IR divergence implies the need for an IR regulator, and predictions of UDW models would then be regulator-dependent. For this reason, a different variant of the UDW model, known as the  \textit{derivative coupling} model, has been used  \cite{Aubry2014derivative,Aubry2018Vaidya,tjoa2020harvesting,Gallock2021harvesting}. This variant couples the detector's monopole moment to the field's \textit{proper time derivative} along the detector's trajectory. This coupling removes the IR ambiguity and has  short-distance behaviour that mimics that of Hadamard states in $(3+1)$ dimensions. Since the effective potential in the radial direction can be neglected in both the near-horizon and asymptotic regimes, the (1+1) model is also accurate quantitatively\footnote{Up to a fixed prefactor $1/2\pi$ due to missing angular direction and the nature of derivative coupling.} in these regimes. This variant has been extensively used to obtain well-controlled results that are expected to be robust in higher dimensions, so long as we do not ask about the physics that depends on the truncated dimensions\footnote{For instance, clearly we cannot study literally the effect of angular momentum of a rotating black hole or the effect of background gravitational waves using two-dimensional truncation.}. \tcb{We  emphasize that in general the derivative coupling model allows us to extract many physical results that agree only qualitatively (but nonetheless are  relevant) with those in (3+1) dimensions, and in some regimes this does give good quantitative agreement. As a recent example in RQI, if we were to study communication between two detectors in a  Schwarzschild background, then the derivative coupling will reproduce the leading-order contribution coming from direct signalling between the detectors, but   will ignore ``indirect'' signalling coming from null ray propagation \textit{around} the black hole \cite{Casals2020communication}.}

In this paper our goal is to extend the applicability of the (1+1)-dimensional derivative coupling model to arbitrary static spherically symmetric (SSS) spacetimes. This is based on the observation that at least for stationary detectors, the constructions used in \cite{Aubry2014derivative,Aubry2018Vaidya,tjoa2020harvesting,Gallock2021harvesting} should allow for this generalization in the same way that models with accelerating mirrors are generalizable to (practically) arbitrary trajectories \cite{cong2019entanglement,Cong2020horizon}. In fact, for SSS geometries we are in a more fortunate situation than that of the accelerating mirror: in principle, one can consider an  arbitrary metric function $f(r)$ from \textit{any theory of gravity}, including those that include higher-curvature corrections to Einstein gravity, such as the recent 4D Einstein-Gauss-Bonnet (EGB) gravity \cite{hennigar2020taking,Fernandes:2020nbq}.   Our work can be viewed as ``maximally'' extending the domain of applicability of the derivative coupling UDW model to include static spherically symmetric backgrounds with horizons in any modified theory of gravity that makes sense in (3+1)-dimensions (before truncation). 

More concretely, in this work we unify the (1+1)-dimensional UDW derivative coupling model for arbitrary $f(r)$ and then exploit this to  calculate detector response on some non-standard metric functions in general relativity:  regular black holes by Hayward \cite{Hayward2006metric}, a ``black bounce'' family of regular black holes, which includes wormhole solutions \cite{Simpson2019wormhole}, and the Schwarzschild-like metric from 4D EGB gravity \cite{hennigar2020taking,Fernandes:2020nbq}. We also show how the notion of thermality in the sense of the  Kubo-Martin-Schwinger (KMS) condition still makes sense in this model, with the KMS temperature precisely given by the familiar formula in black hole thermodynamics (or thermodynamics with horizons), $\mathsf{T}_{\textsc{kms}}=f'(r_\textsc{h})/(4\pi)$, where $r_\textsc{h}$ is the horizon radius \cite{Gibbons1977cosmological,hawking1975particle,Kubiznak2017chemistry}.

Our paper is organized as follows. In Section~\ref{sec: geometry} we introduce the basic description of static spherically symmetric geometries that we need, and some examples that we will use. In Section~\ref{sec: UDW-model} we introduce the derivative coupling UDW model. In Section~\ref{sec: examples} we provide explicit calculations of detector response and analyse how detector sensitivity probes the different metric functions. In Section~\ref{sec: thermality}, we will  formulate a proper notion of thermality with respect to the derivative coupling model, evaluate the detailed-balance condition. We will use the augmented natural units where $c=\hbar=k_B = 1$, and the metric signature is $(-+)$ where a timelike vector $\mathsf{V}$ has negative norm, i.e.,  $g(\mathsf{V},\mathsf{V})<0$. We do not adopt the $G=1$ convention to retain units of length; instead we define $M\coloneqq G\mathsf{M}$, where $\mathsf{M}$ is the ADM mass.

\section{Quantum field theory in static spherically symmetric spacetimes}
\label{sec: geometry}

In this section we review the properties of static spherically symmetric spacetimes and the different coordinate systems adapted to the definitions of some vacuum states in black hole spacetimes. We will then use these to define the truncated two-point Wightman function associated with  what would be the Hartle-Hawking vacuum for arbitrary $f(r)$.

\subsection{Geometry of static spherically symmetric spacetimes}

The most general static spherically symmetric spacetime in $(n+1)$ dimensions has a metric given by the line element
\begin{align}
    \dd s^2 = -f(r)h(r)\dd t^2 + \frac{\dd r^2}{f(r)} + r^2\dd \Omega^2\,,
    \label{eq: SSS-4D}
\end{align}
where $f,h$ are metric functions. For simplicity we restrict to the special case\footnote{There are cases where $h(r)\neq 1$, such as when Einstein gravity couples to matter or when we consider theories like quadratic gravity.} $h(r)=1$. We introduce the tortoise radial coordinates defined by the relation
\begin{align}
    \dd r_*  = \frac{\dd r}{f(r)}\,,
\end{align}
and the double null coordinates $u = t-r_*$ and $v=t+r_*$. Using these, we can rewrite the metric as
\begin{align}
    \dd s^2 = -f(r)\dd u \,\dd v + r^2\dd\Omega^2\,,
\end{align}
where $r=r(u,v)$ is implicitly defined. If we consider metrics where $h(r)\neq 1$, then the tortoise coordinate should be modified by $\dd r_* = \dd r /(f\sqrt{h})$ with the lapse function $h(r)>0$.

The surface gravity of  a body of radius $R$ in a static spherically symmetric spacetime is given by
\begin{align}
    \kappa = \frac{f'(R)}{2}\,,
    \label{eq: surface-gravity}
\end{align}
and we define
\begin{align}
    \kappa_\textsc{h} = \frac{1}{2}f'(r_\textsc{h})
    \label{eq: surface-gravity-horizon}
\end{align}
to be the the surface gravity at the outer horizon of the black hole. Using this, we can define Kruskal coordinates
\begin{align}
    U = -\frac{1}{\kappa_\textsc{h}}e^{-\kappa_\textsc{h} u}\,,\quad V = \frac{1}{\kappa_\textsc{h}}e^{\kappa_\textsc{h} v}\,.
\end{align}
The metric now reads
\begin{align}
    \dd s^2 = -f(r)e^{-2\kappa_\textsc{h}r_*}\dd U\dd V + r^2\dd\Omega^2\,,
\end{align}
where $r = r(U,V)$ and $r_*=r_*(U,V)$ are defined implicitly as a function of $U$ and $V$. It can be checked that this reduces to the usual Kruskal coordinates for more familiar geometries such as Schwarzschild and Reissner-Nordstr\"om black holes. In the case where there is more than one horizon, one could define multiple Kruskal-type coordinates based on their different surface gravities.

\subsection{Klein-Gordon field in static spherically symmetric background}

A real massless scalar field in $(n+1)$-dimensional spacetime conformally coupled to gravity satisfies the covariant Klein-Gordon equation
\begin{align}
    (-\nabla_\mu\nabla^\mu +\xi R)\phi = 0\,,
\end{align}
{where $\xi = \frac{n-1}{4n}$ and} $R$ is the Ricci scalar\footnote{Note that this is only true for test fields. When we demand that the scalar field  backreacts to the background geometry, this is conformal coupling only for Einstein gravity with zero cosmological constant; otherwise linear coupling to the Ricci scalar is not conformally invariant.}. This can be recast into a more convenient form
\begin{align}
    -\frac{1}{\sqrt{-g}}\partial_\mu\rr{\sqrt{-g}g^{\mu\nu}\partial_\nu}\phi +\xi R\phi = 0\,.
\end{align}
The general solution is given by
\begin{align}
    \phi(\sx) = \sum_j a_j u_j(\sx)+a_j^* u_j^*(\sx)\,,
\end{align}
where $u_j$ are eigenmodes of the Klein-Gordon differential operator and $a_j$ are complex numbers. The sum is over discrete or continuous $j$ depending on the background geometry (for example, if the spatial section is compact this sum will be discrete; otherwise it will be a continuous integral over the ``momentum'' variable).

Canonical quantization of the scalar field proceeds by respectively promoting $a_j,a_j^*$ to annihilation and  creation operators $\hat a_j,\hat a_j^\dagger$  that obey the canonical commutation relation 
\begin{align}
    [\hat a_j,\hat a_{j'}^\dagger] = \delta_{jj'} \openone\,,\quad [\hat a_j,\hat a_{j'}] = [\hat a_j^\dagger,\hat a_{j'}^\dagger] = 0\,,
\end{align}
where $\delta_{jj'}$ is either Kronecker or Dirac delta function depending on the discrete/continuous spectrum of the field. This procedure defines an operator-valued distribution $\hat\phi(\sx)$. The mode functions $\{u_j\}$ form an orthonormal basis for the one-particle Hilbert space with inner product furnished by the Klein-Gordon inner product:
\begin{align}
    (f,g)\coloneqq -\ii\int_\Sigma\dd \Sigma^\mu \left(f\nabla_\mu g^* - g^*\nabla_\mu f\right) \,,
\end{align}
where $\Sigma$ is a spacelike Cauchy slice and $\dd\Sigma^\mu$ is the volume form on $\Sigma$. We have
\begin{align}
    (u_j,u_{j'}) = \delta_{jj'}\,,\quad (u_j^*,u_{j'}^*) = -\delta_{jj'} \,,\quad (u_j,u_{j'}^*) = 0\,.
\end{align}
The vacuum state $\ket{0}$ is defined by the state that is annihilated by all $\hat a_j$, i.e., $\hat a_j\ket{0}=0$ for all $j$. Note that in general there are no preferred vacuum states in curved spacetimes even for the static case: for instance, two different vacua associated with  two distinct timelike Killing vectors that are not proportional to one another are in general not (unitarily) equivalent.

Let us now restrict our attention to the two-dimensional truncation of the metric so that by truncating the angular part of Eq.~\eqref{eq: SSS-4D}, the line element now reads
\begin{align}
    \dd s^2 = -f(r)\dd t^2+\frac{\dd r^2}{f(r)}\,.
\end{align}
We can recast this into either Kruskal coordinates or double-null coordinates
\begin{align}
    \dd s^2 &= -f(r) e^{-2\kappa_\textsc{h}r_*}\dd U\dd V = -f(r) \dd u \dd v\,.
\end{align}
It is now clear that the metric is written in conformally flat form: this is always possible because all two-dimensional metrics are conformally flat. Furthermore, in two dimensions the conformal coupling of the scalar field reduces to minimal coupling since $\xi = 0$. 

Since one of our interests  is to study thermalization of detectors, we will consider  Kruskal coordinates $(U,V)$ that are  adapted to the \textit{Hartle-Hawking vacuum} $\ket{0_\textsc{H}}$ of the field. This vacuum state has the property that the mode functions $\{u_j(U,V)\}$ are  positive-frequency eigenmodes with respect to the null generators $\partial_U$ and negative frequency modes $\{u_j^*(U,V)\}$ with respect to $\partial_V$ at the Killing horizon. Using conformal invariance of the field equation, the Klein-Gordon equation reduces to
\begin{align}
    \partial_U\partial_V\phi = 0\,.
\end{align}
The general solution is given by $\phi(U,V) = A(U)+B(V)$ for arbitrary functions $A,B$. Using Fourier mode decomposition, we can write the field operator as
\begin{align}
    \hat\phi(\sx) = \int_0^\infty \frac{\dd \omega }{\sqrt{4\pi\omega}}\bigr[\hat a_\omega e^{-\ii\omega  U} + \hat a_\omega^\dagger e^{\ii\omega  U} 
    + \hat b_\omega e^{-\ii\omega  V} + \hat b_\omega^\dagger e^{\ii\omega  V}\bigr]\,.
    \label{eq: fourier-mode}
\end{align}
The first two terms define the left-moving modes and the last two terms define the right-moving modes. The Fourier mode decomposition \eqref{eq: fourier-mode} should be understood as having an infrared (IR) regulator $\Lambda>0$, since massless scalar fields in non-compact two-dimensional spacetimes are generically IR-divergent. 

The Wightman two-point function for the scalar field with respect to the Hartle-Hawking vacuum is thus given by
\begin{align}
    W_\textsc{h}(\sx,\sx') &= \int_\Lambda^\infty \frac{\dd\omega\dd\omega'}{4\pi\sqrt{\omega\omega'}} \braket{0_\textsc{h}| a_\omega a_{\omega'}^\dagger+  b_\omega b_{\omega'}^\dagger|0_\textsc{h}} \notag\\
    &= -\frac{1}{4\pi}\log\bigr[-\Lambda^2(\Delta U-\ii\epsilon)(\Delta V - \ii\epsilon)\bigr]\,, 
    \label{eq: wightman-hhi}
\end{align}
where $\Lambda$ is the IR regulator. It is this IR divergence that the derivative coupling model attempts to get rid of, in addition to the ``wrong'' short-distance scaling. 

Our main result is based on the observation that since the vacuum state is completely specified by the two-point function, we could take the two-point function as the definition of the Hartle-Hawking vacuum. However unlike the standard calculation, the $(U,V)$ coordinates are associated with the Kruskal coordinates of an \textit{arbitrary static spherically symmetric geometry}.  We will see in the next section that within the  derivative coupling particle detector model, the generality of the metric function $f(r)$ will \textit{not} pose significantly more difficulty for   calculating the response of a detector.

Finally, we remark that we could have defined the other two standard vacua --- Boulware and Unruh vacua $\ket{0_\textsc{B}},\ket{0_\textsc{U}}$ --- by following similar procedures. The details are discussed in \cite{Aubry2018Vaidya,tjoa2020harvesting,birrell1984quantum}, but we quote here the Wightman functions associated with these two vacua:
\begin{align}
    W_\textsc{B}(\sx,\sx') &= -\frac{1}{4\pi}\log\bigr[-\Lambda^2(\Delta u-\ii\epsilon)(\Delta v - \ii\epsilon)\bigr]\,,
    \label{eq: wightman-boulware}\\
    W_\textsc{U}(\sx,\sx') &= -\frac{1}{4\pi}\log\bigr[-\Lambda^2(\Delta U-\ii\epsilon)(\Delta v - \ii\epsilon)\bigr]
    \label{eq: wightman-unruh}\,.
\end{align}
Although the Boulware vacuum has divergent stress-energy tensor at the Killing horizon,  it has been shown to be   relevant at the  early-time/far from the black hole limit of the vacuum state associated with collapsing matter, such as a Vaidya background \cite{Aubry2018Vaidya,tjoa2020harvesting}. The Unruh vacuum is constructed to mimic the late-time regime of an evaporating black hole, essentially by replacing the ingoing Hartle-Hawking mode with the ingoing Boulware modes. Thus  the Unruh vacuum   represents a  non-equilibrium situation associated with outward thermal flux.

\section{Derivative coupling Unruh-DeWitt model}
\label{sec: UDW-model}

In this section we review the basics of the Unruh-DeWitt detector model for the   entanglement harvesting protocol. Although not the original Unruh-DeWitt model (hence sometimes said to be `UDW-like'), we will call the derivative-coupling version an Unruh-DeWitt model as well for convenience.

\subsection{Time evolution in derivative coupling UDW model}

The derivative coupling UDW model is defined as a pointlike two-level quantum system (a qubit) which interacts with a scalar field via the following interaction Hamiltonian (in the interaction picture):
\begin{align}
    \H_I(\tau) &= \lambda \chi (\tau) \hat{\mu}(\tau) \otimes u^\mu \nabla_\mu  \hat\phi(\sx(\tau))\,,
\end{align}
where $u^\mu$ is the 4-velocity of the detector parametrized by proper time $\tau$; $\lambda$ denotes the coupling strength and $\hat\mu$ is the monopole moment, given in terms of the detector proper time:
\begin{align}
    \hat\mu(\tau) = \hat\sigma^+ e^{\ii\Omega \tau} + \hat\sigma^- e^{-\ii\Omega \tau}\,,
\end{align}
with $\Omega$ the detector gap, $\hat\sigma^\pm$ the ladder operators of an  $\mathfrak{su}(2)$ algebra, and $\chi(\tau)$ is the switching function that controls the duration of interaction.  For simplicity we will consider Gaussian switching functions 
\begin{align}
    \chi(\tau)=e^{-\frac{\tau^2}{T^2}}\,,
    \label{eq: switching}
\end{align}
where $T$ prescribes the duration of interaction. Due to time translation symmetry, we can align $\tau=0$ to be equal to $t=0$. 

The time evolution operator for the detector-field system is given by 
\begin{align}
    \hat U = \mathcal{T}\exp\rr{-\ii\int \dd \tau \H_I(\tau)}\,, 
\end{align}
where $\mathcal{T}$ is the time-ordering operator.  In the weak coupling regime, we can perform a Dyson series expansion 
\begin{align}
    \hat U = \openone + \hat U^{(1)}+ \hat U^{(2)} + O(\lambda^3)\,,
    \label{Dyson}
\end{align}
whose first two terms are
\begin{subequations}
\begin{align}
    \hat U^{(1)} &= -\ii\int_{-\infty}^\infty \dd \tau \,\H_I(t)\,,\\
    \hat U^{(2)} &= -\int_{-\infty}^\infty\dd \tau \int^\tau_{-\infty}\dd \tau' \,\hat H_I(\tau )\hat H_I(\tau')
\end{align}
\end{subequations}
and where $\hat U^{(k)}$ is of order $\lambda^k$. We take the initial state to be the uncorrelated state  
\begin{align}
    \rho_0 = \ket{g}\!\bra{g}\otimes\ket{0_\alpha}\!\bra{0_\alpha}\,,
    \label{eq: initial-state}
\end{align}
where $\alpha=\textsc{B,U,H}$ (also $\alpha=\textsc{V,S}$ if we include the Vaidya and static star; see the discussion at the end of the section) labels the vacuum state of the field and
$\ket{g},\ket{e}$ are the ground and excited states of the free Hamiltonian $\mathfrak{h} = \frac{\Omega}{2}(\hat\sigma_z+\openone)$. These states are related by the $\mathfrak{su}(2)$ ladder operators so that $\hat \sigma^+\ket{g} = \ket{e}$ and  $\hat \sigma^-\ket{e} = \ket{g}$. 

Since we are interested in the final state of the detector, we will take the  partial trace over the field's degrees of freedom after the unitary time evolution. That is, the final state of the detector is given by
\begin{align}
    \rho_{\textsc{d}} = \tr_\phi\left[\hat U\rho_{0}\hat U^\dagger\right]\,.
\end{align}
Using the Dyson series expansion \eqref{Dyson}, the final state of the detector can be written as a perturbative expansion 
\begin{align}
    \rho_\textsc{d} &= \rho_{\textsc{d},0} + \rho^{(1)} + \rho^{(2)} + O(\lambda^3)\,,
\end{align}
where $\rho^{(k)}$ is of order $\lambda^k$:
\begin{subequations}
\begin{align}
    \rho^{(1)} &= \tr_\phi\left[\hat U^{(1)}\rho_0 + \rho_0 U^{(1)\dagger}\right]\,,\\
    \rho^{(2)} &= \tr_\phi\left[\hat U^{(1)}\rho_0\hat U^{(1)\dagger} +  \hat U^{(2)}\rho_0 + \rho_0 U^{(2)\dagger}\right]\,.
\end{align}
\end{subequations}
The choice of initial state in Eq.~\eqref{eq: initial-state} implies that $\rho^{(1)} = 0$ since the one-point function $\braket{0|\hat\phi(\sx)|0} = 0$ for all $\sx$. Therefore, the leading order contribution in perturbation theory is $\rho^{(2)}$. In the ordered basis $\{\ket{g},\ket{e}\}$, one can show that the matrix representation of the final state reads
\begin{align}
    \rho_\textsc{d} = \begin{pmatrix}
    1-P & 0 \\ 0 & P
    \end{pmatrix} + O(\lambda^4)\,,
\end{align}
where $P\equiv P(\Omega)$ is the excitation probability of the detector: 
\begin{align}
    P(\Omega) &= \lambda^2\int \dd\tau \int \dd\tau'\, \chi(\tau)\chi(\tau')e^{-\ii\Omega(\tau-\tau')}\mathcal{A}_\alpha(\tau,\tau')\,.
    \label{eq: local-noise}
\end{align}
The bi-distribution $\mathcal{A}_\alpha(\tau,\tau')\equiv \mathcal{A}_\alpha(\sx(\tau),\sx(\tau'))$ is the proper-time derivative of the vacuum Wightman function along the detector's trajectory:
\begin{align}
    \mathcal{A}(\tau,\tau') =  \partial_\tau\partial_{\tau'}W_\alpha(\sx(\tau),\sx(\tau'))\,,\quad W_\alpha(\sx,\sx') = \braket{0_\alpha|\hat\phi(\sx)\hat{\phi}(\sx')|0_\alpha}\,.
    \label{eq: derivative-Wightman}
\end{align}
We remark that for $\Omega>0$, $P(-\Omega)$ corresponds to \textit{de-excitation probability} from excited to ground state.

\subsection{Derivative coupling Wightman two-point distributions for static spherically symmetric spacetimes}

The remaining task is to calculate the derivative-coupling Wightman function for different vacua of interest. Let us define the shorthand $\mathcal{A}_\alpha(\tau,\tau') \equiv \mathcal{A}_\alpha(\sx(\tau),\sx(\tau'))$, where $\alpha=B,U,H$. We also use write $\dot{y}\equiv \pd_\tau [y(\tau)]$, and $\dot{y}'\equiv \pd_{\tau'}[y(\tau')]$. Taking a proper-time derivative of Eqs.~\eqref{eq: wightman-boulware},\eqref{eq: wightman-unruh}, and \eqref{eq: wightman-hhi}, we obtain (cf. \cite{Aubry2014derivative,Aubry2018Vaidya,tjoa2020harvesting,Gallock2021harvesting})
\begin{subequations}
\begin{align}
    \mathcal{A}_\textsc{B}(\tau,\tau') 
    &= -\frac{1}{4\pi}\left[\frac{\dot{u}\dot{u}'}{(u-u'-\ii\epsilon)^2}+\frac{\dot{v}\dot{v}'}{(v-v'-\ii\epsilon)^2}\right]\,,
    \label{eq: derivative-boulware}\\
    \mathcal{A}_\textsc{U}(\tau,\tau') 
    &= -\frac{1}{4\pi}\left[\frac{\dot{U}\dot{U}'}{(U-U'-\ii\epsilon)^2}+\frac{\dot{v}\dot{v}'}{(v-v'-\ii\epsilon)^2}\right]\,,
    \label{eq: derivative-unruh}\\
    \mathcal{A}_\textsc{H}(\tau,\tau') 
    &= -\frac{1}{4\pi}\left[\frac{\dot{U}\dot{U}'}{(U-U'-\ii\epsilon)^2}+\frac{\dot{V}\dot{V}'}{(V-V'-\ii\epsilon)^2}\right] \,.
    \label{eq: derivative-hhi}
\end{align}
\end{subequations}
Notice that the IR cutoff has dropped out of the two-point distributions, and $\mathcal{A}_\alpha(\tau,\tau')$ has the   power-law decay behaviour expected for the vacuum Wightman functions of a massless scalar field in (3+1) dimensions. We emphasize that while the \textit{form} of the derivative coupling two-point distribution is the same as that of two-dimensional Schwarzschild case, the expressions above are valid for \textit{arbitrary static spherically symmetric geometries}. This is made possible by the conformal invariance of the massless wave equation and conformal flatness in two dimensions. 

One important aspect that we should emphasize is the role of boundary conditions on the field. For example, we could also consider the derivative coupling Wightman function associated with the field $\phi$ subject to Dirichlet boundary conditions at $r=0$, such as when one considers collapsing shell geometries such as Vaidya spacetimes within Einstein gravity \cite{Aubry2014derivative,Aubry2018Vaidya}. This is also relevant when we consider a quantum field living on a background spacetime with a static spherically symmetric star, where the Boulware vacuum is expected to be a good approximation of the exterior vacuum state (see, e.g., \cite{Visser1996polarization}). In these cases, the Wightman function acquires a term that comes from the Dirichlet boundary, analogous to the accelerating mirror case. Following procedures similar to the situations considered in  \cite{birrell1984quantum,Aubry2014derivative,Aubry2018Vaidya, tjoa2020harvesting, cong2019entanglement,Cong2020horizon}  we get
\begin{align}
    W_\textsc{V}(\sx,\sx') &= -\frac{1}{4\pi}\log\left[\frac{(\bar{U}-\bar{U}'-\ii\epsilon)(v-v'-\ii\epsilon)}{(\bar{U}-v'-\ii\epsilon)(v-\bar{U}'-\ii\epsilon)}\right]\,,
    \label{eq: derivative-vaidya}\\
    W_\textsc{S}(\sx,\sx') &= -\frac{1}{4\pi}\log\left[\frac{(u-u'-\ii\epsilon)(v-v'-\ii\epsilon)}{(u-v'-\ii\epsilon)(v-u'-\ii\epsilon)}\right]\,,
    \label{eq: derivative-star}
\end{align}
where $\bar{U} = -4M(1+\textsc{W}(-4MU/e))$, and $\textsc{W}(z)$ is the Lambert-W function. The Wightman functions are given by $W_\textsc{V}$ and $W_\textsc{S}$ are respectively associated with  the Vaidya vacuum and the Boulware vacuum for the star's exterior\footnote{Observe that $W_\textsc{S}(\sx,\sx')$ has exactly the same form as the Wightman function for a  \textit{static mirror} at the origin in two dimensions. The difference lies in the radial coordinates, since for the star it will be the tortoise radial coordinates $r_\star(r)$ instead of $r$ and there is gravitational redshift $\dd t/\dd\tau\neq 1$. }. Note that in this case, there is no longer an  IR divergence but the leading short-distance behaviour does not scale with power-law behaviour like a scalar field in (3+1) dimensions. The Wightman functions in the derivative coupling version is then given by Eq.~\eqref{eq: derivative-Wightman}.

Another situation where the boundary condition particularly matters is when the background geometry is not globally hyperbolic, with QFT in Anti-de Sitter (AdS) spacetime being the most well-studied of all \cite{Isham1978AdS}. The boundary condition is needed at the conformal boundary when one works with the universal covering space, and for massless scalar fields there are several boundary conditions that work (see, e.g, \cite{Isham1978AdS,Ana2021time}). In this case, the boundary conditions will be captured by the Wightman function $W(\sx,\sx')$ with respect to the truncated metric, similar to the Vaidya and star scenarios, and then one takes the proper time derivative on both arguments. Some aspects of QFT on AdS$_2$ and its topological identification (e.g., those that produce time machines) were recently investigated  \cite{Pitelli2021UDW,Pitelli2019AdS2,Pitelli2019boundaryAdS,Dappiaggi2016AdS, emparan2021holography,Ana2021time} and they are readily extended to the derivative coupling variant \cite{Ana2021timemachine2}.

\section{Detector responses}
\label{sec: examples}

What we have done so far is to show that the derivative coupling model can be generalized to an arbitrary SSS metric \eqref{eq: SSS-4D}, with or without horizons, whose truncation allows us to work out the various Wightman functions of interest. The truncation allows us to explore some of the physics that is expected to remain robust, despite the absence of the angular direction, such as the detailed balance condition or how the stationary detector response varies with the coupling constants of the gravitational theory of interest (thermalization will be discussed in Section~\ref{sec: thermality}). In this case, the physical input associated with the background geometry is given by the following:
\begin{enumerate}
    \item The coupling constants of the gravitational theory, e.g., the cosmological constant or higher curvature coupling.
    
    \item The parameters of the metric function $f(r)$, such as mass, charge, or some extra length scales (e.g., in the case of regular static Hayward black holes).
    
    \item The choice of vacuum states of the test quantum field. For instance, if we consider the Vaidya metric in Einstein gravity, then the 2D truncation (considered in \cite{Aubry2014derivative,tjoa2020harvesting}) will require the reflecting boundary at $r=0$, and similarly for exterior of static stars; in contrast, eternal Schwarzschild black holes admit three Hadamard states (Boulware, Unruh, Hartle-Hawking) at the exterior.
\end{enumerate}
Procedurally, one first decides on the gravitational theory and the corresponding metric function $f(r)$ that the theory admits. Next, one checks the global structure of the underlying spacetime (before truncation) and the boundary conditions (or lack thereof). Once these are fixed, the Wightman function in the truncated spacetime can be derived from standard canonical quantization, and the derivative coupling Wightman function is obtained by proper time derivatives.

\subsection{Concrete examples of static spherically symmetric spacetimes}

Since this generalization opens up a lot of choices for the static spherically metrics in (3+1)-dimensional geometries, we will choose some representative (simple) examples to illustrate the utility. We will consider four representative examples:
\begin{enumerate}[leftmargin=*,label=(\alph*)]
    \item Reissner-Nordstr\"om-(A)dS black hole in Einstein gravity, from which we take the Schwarzschild family as a baseline for comparison;
    \item Hayward black hole, the simplest (early) example of a regular black hole with no curvature singularity \cite{Hayward2006metric};
    \item Static spherically symmetric black hole in (3+1)-dimensional Einstein-Gauss-Bonnet (EGB) gravity \cite{hennigar2020taking,Fernandes:2020nbq}, an example from a (ghost-free) modified theory of gravity;
    \item ``Black bounce'' metrics, a family of metric functions $f(r)$ that interpolate between Schwarzschild solution and (traversable) wormhole geometries \cite{Simpson2019wormhole}.
\end{enumerate}
These examples are simple enough to include a one-parameter family of modifications of Schwarzschild geometries that allow us to make clear comparisons of what we can expect. Below we include brief descriptions of each example before we proceed to compare the detector responses in the derivative coupling UDW model.

\subsubsection{Einstein gravity} 

For black holes in (3+1)-dimensional Einstein gravity with/without cosmological constant, the most general static spherically symmetric metric is given by the Reissner-Nordstr\"om-(Anti-)de Sitter (RN-(A)dS) metric
\begin{align}
    f(r) = 1-\frac{2M}{r}+\frac{Q^2}{r^2} \pm \frac{r^2}{L^2}\,,
    \label{eq: RN-AdS}
\end{align}
where the cosmological constant is related to the (A)dS length by $\Lambda = \pm 3/L^2$, $M$ and $Q$ are the ADM mass and charge of the black hole. This metric arises from the Einstein-Hilbert action coupled to the electromagnetic field:
\begin{align}
    S_{EH} = -\frac{1}{16\pi G}\int \dd^4\sx \sqrt{-g}\,\rr{R-2\Lambda -4\pi G F_{\mu\nu}F^{\mu\nu}}\,.
\end{align}
Focusing on Hartle-Hawking state, the relevant calculation is to obtain the correct Kruskal coordinates $U,V$ for these black holes.

 Since the representative examples we study are uncharged\footnote{We remark that the detector response calculations (or rather, transition rate calculations) for Reissner-Nordstr\"om black holes were recently investigated in order to understand the effect of Cauchy horizons on infalling detectors \cite{juarezaubry2021quantum}. }, we will take as a baseline the Schwarzschild black hole in Einstein gravity case to the standard Schwarzschild metric with $f(r) = 1-2M/r$.
Furthermore, quantum field theory in (A)dS$_4$ is very well-studied (see \cite{birrell1984quantum} for dS$_4$ and \cite{Isham1978AdS} for AdS$_4$). There are also recent $(1+1)$-dimensional examples in AdS$_2$ \cite{Pitelli2021UDW} and time machine geometry locally isometric to AdS$_2$ \cite{Ana2021time,Ana2021timemachine2}. These latter cases are amenable to derivative coupling modification, but we will not pursue these further  since their $(3+1)$-dimensional counterparts (without derivative coupling) are already computationally tractable.

\subsubsection{Hayward metric}

A regular black hole, one that does not possess a curvature singularity, was first studied by Bardeen \cite{Bardeen1968regular}, where the black holes are required to satisfy reasonable conditions such as the correct metric fall-off at infinity and an Einstein tensor obeying weak energy conditions. Regular black holes have been extensively studied in the literature, and we point the reader to \cite{Frolov2016regular,ansoldi2008spherical} and references therein for more details and refinements. For our purposes we only need to demonstrate our calculation using a simple example of a regular black hole metric, namely the one provided by Hayward \cite{Hayward2006metric}.

The Hayward black hole is arguably the simplest minimal model of regular black holes in (3+1)-dimensional Einstein gravity, with metric function given by
\begin{align}
    f(r) = 1-\frac{2Mr^2}{r^3+2l^2M}\,,
\end{align}
where $l$ is some fixed length scale that needs to be chosen \textit{a priori}. This length scale $l$ can be much larger than the Planck scale itself but not smaller, since this metric is expected to be only valid at most in the semi-classical regime where gravity can still be treated classically. This metric has the property that near the core the metric approaches the de Sitter limit $f(r)\sim 1-r^2/l^2$ with $l$ taking the role of the de Sitter radius; far away the metric approaches the standard Schwarzschild metric with $f(r)\sim 1-2M/r$. There exists a critical mass
\begin{align}
    m_c = \frac{3\sqrt{3}}{4}l\,,
    \label{eq: critical-hayward}
\end{align}
such that $m\geq m_c$ has outer and inner horizons $r_\pm$ (like Reissner-Nordstr\"om and its extremal limit), and for our purposes we take the detector to be at $r>r_+$ (i.e., $r_H = r_+$). For $m<m_c$, the spacetime contains no black hole but the regular core means there is no naked singularity, unlike in the case of an ``overcharged'' Reissner-Nordstr\"om metric. In the limit $m\gg m_c$, the outer horizon is approximately the Schwarzschild radius $r_+\approx 2M$ and the inner horizon is approximately $r_-\approx l$ (this ``core'' is gravitationally repulsive since the geometry near the core is de Sitter). 

For our purposes in comparing various static spherically black hole geometries with horizons, we will restrict our attention to the case where $m>m_c$. This is also where the (generalized) Hartle-Hawking state in the previous section is most relevant. However, we emphasize that as far as the UDW model on the (truncated) static spherically symmetric background is concerned, nothing prevents us from considering $m<m_c$ where there is no black hole. In this case, if we regard the de Sitter core as a hard core with Dirichlet boundary condition, then  the Wightman function that we need to consider for the scalar field  is ``Boulware-like'',  analogous to \eqref{eq: derivative-star} for static star's exterior geometry with appropriate modification to the tortoise radial coordinate.

\subsubsection{Einstein-Gauss-Bonnet gravity}

The four-dimensional Einstein-Gauss-Bonnet (EGB) gravity has attracted much interest in recent years as a modified theory of gravity that includes higher-curvature corrections to Einstein gravity. This is because in four dimensions, it was thought that general relativity is the unique gravitational theory whose Lagrangian yields second order equations of motion for the metric. Many higher-curvature modifications break this property, which most often leads to an ill-posed initial value problem or the existence of ghost degrees of freedom. Four-dimensional EGB gravity turns out to have this attractive property\footnote{The property that the metric equation of motion is second-order is shared by a well-known class of theories known as \textit{Lovelock gravity} \cite{Lovelock1971}, but the higher-curvature corrections only modify the equation of motion non-trivially in five dimensions or higher.}, however its construction is somewhat subtle. The first construction given in \cite{Glavan2020EGB} suffers from dimension-dependent limiting procedures; a more rigorous, ``intrinsic'' construction of 4D EGB theory without relying on a  higher-dimensional ``Kaluza-Klein-type'' compactification procedure  \cite{hennigar2020taking,Fernandes:2020nbq}. 
The readers are invited to check \cite{hennigar2020taking} and references therein for the general theory and its applications.

The static spherically symmetric Schwarzschild-(A)dS-like metric in four-dimensional EGB gravity is given by two ``branches'' \cite{hennigar2020taking} (also derived in \cite{Lu2020horndeski} via the compactification techniques used in \cite{Glavan2020EGB}):
\begin{align}
    f_\pm (r) = 1+\frac{r^2}{2\alpha}\rr{1\pm \sqrt{1+\frac{4}{3}\alpha\Lambda+\frac{8\alpha M}{r^3}}}\,,
    \label{eq: EGB-metric}
\end{align}
where $\alpha$ is the (reduced) Gauss-Bonnet coupling\footnote{It is not quite the actual Gauss-Bonnet coupling, since the original Gauss-Bonnet gravity is only non-trivial in five dimensions or higher ($n\geq 4$) \cite{hennigar2020taking}.}. Only the $f_-(r)$ branch approaches the Schwarzschild-(A)dS metric when $\alpha\to 0$ and has the right $\Lambda$-dependent asymptotic fall-offs. The $f_+(r)$ branch represents a black hole that has a  curvature singularity but is not asymptotically flat for all choices of parameters $\alpha,\Lambda$.

For simplicity we will restrict our attention to the asymptotically flat case $(\Lambda=0)$. Then the $f_-(r)$ solution quickly approaches the Schwarzschild metric at large $r$, and it modifies the horizon so that now we have outer and inner horizons $r_{\pm}= M \pm \sqrt{M^2-\alpha}$. This simplification allows transparent comparison with the Schwarzschild case. If we wish to consider, for instance, $\Lambda<0$, one should take note of the fact that since the geometry is asymptotically AdS, there is a need to impose boundary conditions at the conformal (timelike) boundary. Therefore, the (derivative) Wightman function one needs to consider is not the one in \eqref{eq: derivative-hhi} but includes extra terms due to boundary conditions, analogous to \eqref{eq: derivative-vaidya} or \eqref{eq: derivative-star}. So long as one takes care of these extra conditions (outlined at the beginning of Section~\ref{sec: examples} and at the end of Section~\ref{sec: UDW-model}), {it is possible to consider the asymptotically AdS case.}

We make a parenthetical remark that unlike the  Schwarzschild geometry, the metric function \eqref{eq: EGB-metric} has \textit{finite} limiting value as $r\to 0$ but there is still curvature singularity: the leading behaviour of the Kretschmann scalar invariant $K=R^{\mu\nu\rho\sigma}R_{\mu\nu\rho\sigma}$ at the origin is controlled by $f''_-(r)\propto r^{-4}$. 

\subsubsection{Black bounce}

In \cite{Simpson2019wormhole} a family of static spherically symmetric metrics was proposed, which reads
\begin{align}
    \dd s^2 = -\rr{1-\frac{2M}{\sqrt{r^2+a^2}}}\dd t^2+\frac{\dd r^2}{1-\frac{2M}{\sqrt{r^2+a^2}}}+(r^2+a^2)\dd\Omega^2\,,
\end{align}
where $a\geq 0$. Therefore we have yet another ``minimal'' modification of the Schwarzschild geometry (with the Schwarzschild spacetime recovered at $a=0$), but the angular part acquires a correction to the radial coordinate (unlike the Hayward or EGB metrics). For our purposes, what we need to be aware of (see \cite{Simpson2019wormhole} and references therein for more details) is the fact that this solution represents different objects for different choices of $a$:
\begin{enumerate}[label=(\roman*),leftmargin=*]
    \item $a>2M$: this represents (two-way) \textit{traversable wormholes};
    \item $a=2M$: this represents a (one-way) wormhole with null throat at $r=0$;
    \item $0<a<2M$: this represents a  \textit{regular} black hole, with finite curvature as well as scalar curvature invariants;
    \item $a=0$: the Schwarzschild black hole.
\end{enumerate}
What we will be interested in is case (iii). In \cite{Simpson2019wormhole} the metric for $a\in (0,2M)$ is called a ``black bounce'' (BB) geometry because the maximal analytic extension across $r=0$ allows for infinitely many copies of the same universe into the future/past (see Figure 4 of \cite{Simpson2019wormhole}). Notice that within our UDW model, since we are going to truncate the angular part, the black bounce metric reduces to choosing the metric function
\begin{align}
    f(r) = 1-\frac{2M}{\sqrt{r^2+a^2}}\,, 
\end{align}
with the implicit understanding that for $a>0$ the constant-$r$ slices are spheres of radius $\sqrt{r^2+a^2}$. We will call this the black bounce (BB) metric for convenience.

Again, as per the discussions earlier, in principle nothing prevents us from studying cases (i) and (ii). One just needs to be aware of the extra conditions that one may need to check before choosing the correct Wightman function. For instance, in case (i) the traversable wormhole (of minimum radius $r=a$) has no horizon, so the Boulware-like vacuum analogous to \eqref{eq: derivative-boulware} (again, with suitable modification to the tortoise coordinates) is both natural and tenable, unlike the Schwarzschild case.

\subsection{Comparison of detector responses}

We are now ready to compare detector responses for different representatives of SSS metrics. We will be comparing Hayward, EGB, and BB metrics with respect to the Schwarzschild case, so instead of computing detector responses we will use the normalized ratio
\begin{align}
    \mathcal{R}\coloneqq \frac{P_\text{mod}}{P_\text{Schw}} - 1\,,
\end{align}
where $P_{\text{mod}}$ represents the transition probability (detector response) when the metric function is one of Hayward, EGB, or BB; $P_{\text{Schw}}$ represents the transition probability for the Schwarzschild metric. The detector responses will be computed with respect to Hartle-Hawking vacuum, so we will be using the derivative Wightman function \eqref{eq: derivative-hhi}.

In our calculation we will mostly measure quantities in units of the switching duration $T$, which we will not vary (unless otherwise stated). The calculations will be done numerically using the method outlined in \cite{tjoa2020harvesting}, so we will take the proper distance $d_\text{det} \coloneqq d(r_H,r_\text{det})$ (in units of $T$) of the detector from the horizon to be
\begin{align}
    d \coloneqq \frac{d_\text{det}}{T} \geq b\,,\quad b=\frac{1}{25}\,,
\end{align}
where $d$ is the dimensionless proper distance and $d=b$ is the closest approach in our calculation\footnote{Numerically, we can take $b$ to be arbitrarily small,  in which case 
more computation time may be required and numerical stability issues may arise at
$d\approx b$ when $b$ is smaller.}.

\begin{figure}[tp]
    \centering
    \includegraphics[scale=0.77]{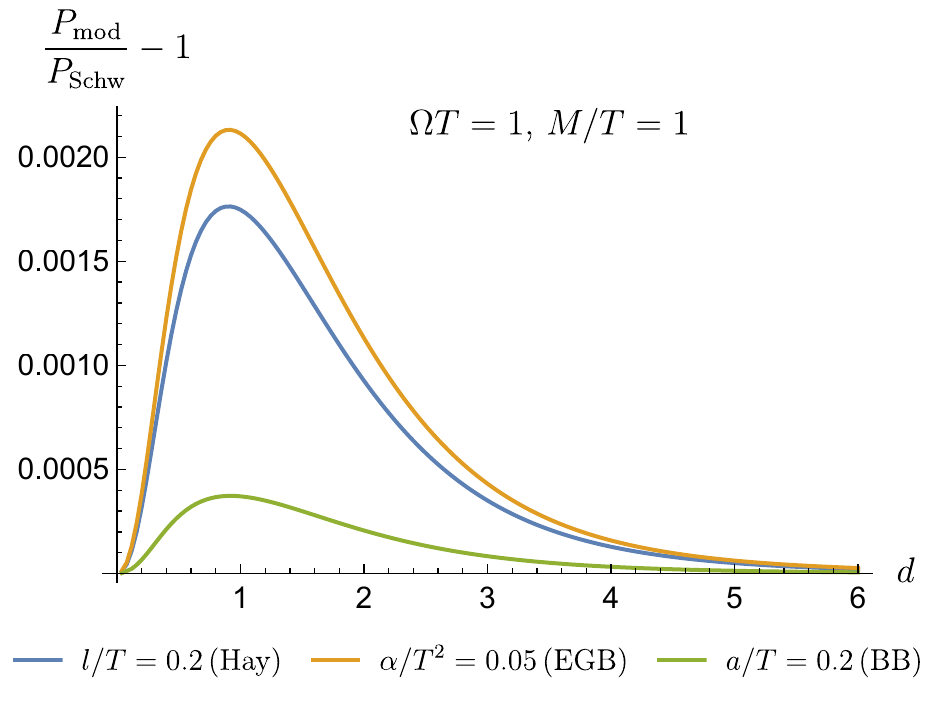}
    \includegraphics[scale=0.77]{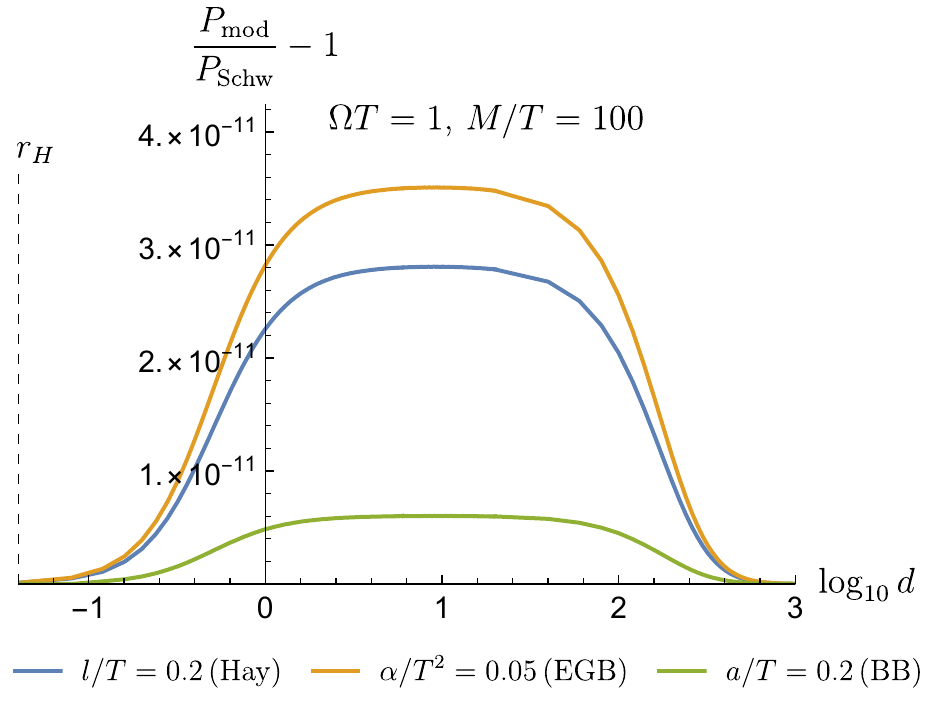}
    \caption{The probability differences, computed as the ratio $\mathcal{R}=P_{\text{mod}}/P_{\text{Schw}}-1$ as a function of (dimensionless) proper distance  $d$. We set $\Omega T = 1$ and deformation parameter $l/T = 0.2$ for Hayward, $\alpha/T^2=0.05$ for EGB and $a/T = 0.2$ for BB. (a) Small mass regime $M/T=1$. (b) Large mass regime $M/T = 100$. In both cases we see that at intermediate distance away from the black hole the differences between the metrics are maximal, and the range of distances for which this is observable increases with distance.}
    \label{fig: Fig1-radial}
\end{figure}

We are now ready to compare the detector responses between various metric functions. In Figure~\ref{fig: Fig1-radial} we plot the probability ratio $\mathcal{R}$ as a function of proper distance from the horizon. We pick  representative deformation parameters $l,\alpha,a$ (for Hayward, EGB and BB respectively), considering the small mass regime   in Figure~\ref{fig: Fig1-radial}(a) and the  large mass regime  in Figure~\ref{fig: Fig1-radial}(b). We see that in all cases, the deformation parameter leads to larger detector responses\footnote{The exception here is EGB case when $\alpha<0$, in which case one can show that the detector response is \textit{lower} than Schwarzschild case. We will not study this further in this work.}, with the largest difference occuring at sufficiently large but finite proper distance away from the horizon. This is because, regardless of the deformation parameters and the theory of gravity from which the metric is obtained, all the metric functions considered here are rational functions of $r$ with common property  that $f(r)\to 0$ as $r\to r_H$ and $f(r)\to 1$ as $r\to \infty$. Therefore, the dominant differences between the detector responses must appear at some  ``intermediate distance'' from the black hole. Note that larger mass black holes suppress the difference but increase the range of distances at which the difference is {maximized}: in Figure~\ref{fig: Fig1-radial}(a) significant probability differences are confined mostly at $d\approx 1$, while in Figure~\ref{fig: Fig1-radial}(b) significant probability differences occur all the way up to $d\approx 100$ (plotted on a  logarithmic scale).

The second observation from Figure~\ref{fig: Fig1-radial} is that although the metric functions may come from different theories of gravity and the deformation parameters serve to describe very different types of black holes, there are regimes where the detector responses are not sensitive enough to detect the differences. To see this, if we take the small-$l$, small-$\alpha$ and small-$a$ expansion of the three metrics, we get
\begin{equation}
\begin{aligned}
    f_\text{Hay}(r) &\approx 1-\frac{2M}{r} + \frac{4M^2l^2}{r^4} + O(l^2) \,,\\
    f_\text{EGB}(r) &\approx 1-\frac{2M}{r} + \frac{4M^2\alpha}{r^4} + O(\alpha^2)\,,\\
    f_\text{BB}(r) &\approx 1-\frac{2M}{r} + \frac{M^2a}{r^3} + O(a^2)\,.
    \label{eq: small-expansion}
\end{aligned}
\end{equation}
The smallness of the {subleading corrections} can be measured relative to $M$ (or Schwarzschild radius $r_\textsc{h}=2M$) and is valid for all $r>r_\textsc{h}$. \{Observe from Eq.~\eqref{eq: small-expansion} that  if $\alpha = l^2 $ and $l\ll 2M$ then $f_\text{EGB}$
  \textit{indistinguishable}
from $f_\text{Hay} $  at leading order both classically and quantum mechanically (via detector response). In contrast, because of the $r^{-3}$ fall-off behaviour for the BB metric, there is no way to tune $a$  so that the detector response is indistinguishable in principle \textit{everywhere} from EGB and Hayward cases. Furthermore, since the $O(r^{-3})$ and $O(r^{-4})$ fall-off behaviour in Eq.~\eqref{eq: small-expansion} are not present in the standard RN-(A)dS solution \eqref{eq: RN-AdS}, we have here an explicit example where the UDW model is useful for probing non-standard geometries as well as gravitational theories beyond general relativity. 

\tcb{We should note in passing that one could also compare the probability differences between different metrics at the same gravitational redshift instead of the same proper distance from the black hole horizon. In this case, it can be checked that the larger the redshift   (the closer to the black hole), the larger the differences from a Schwarzschild background. This can be attributed to the fact that all metric functions have the same asymptotic behaviour at large radii (dominated by the leading $1/r$ part) but their near-horizon expansions are different depending on the deformation parameters.}

\begin{figure}[tp]
    \centering
    \includegraphics[scale=0.8]{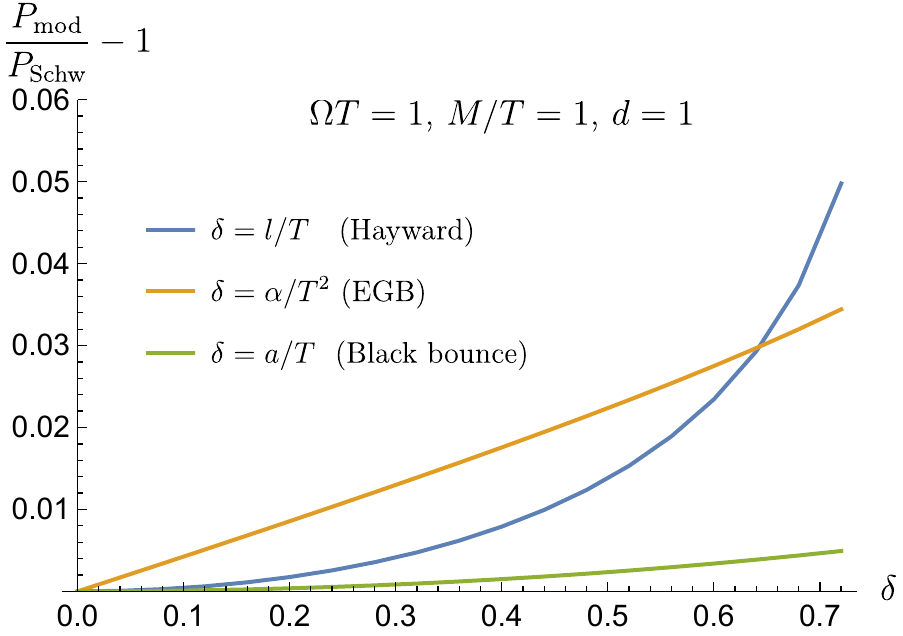}
    \includegraphics[scale=0.8]{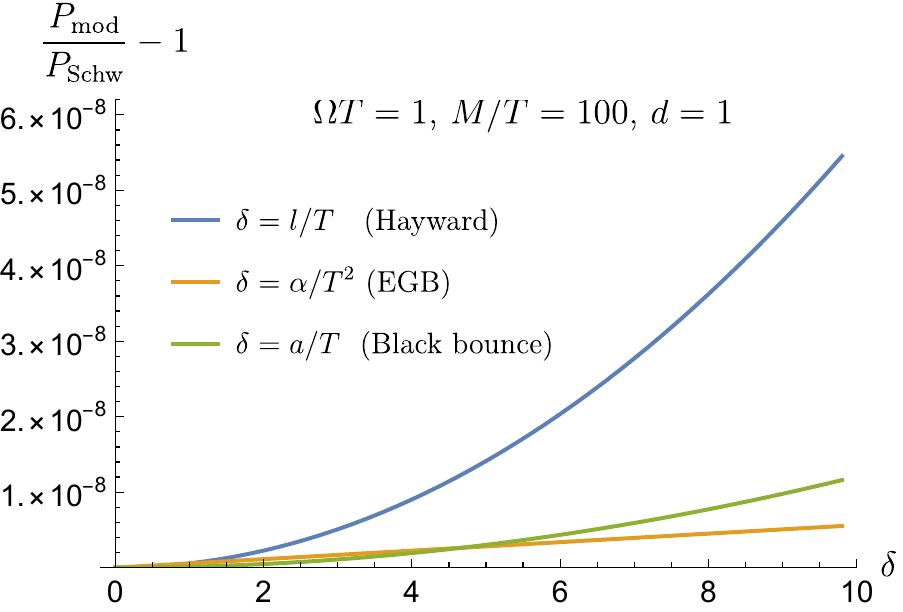}
    \caption{The probability differences, computed as ratio $\mathcal{R}=P_{\text{mod}}/P_{\text{Schw}}-1$ as a function of (dimensionless) deformation parameter $\delta$ as defined in Eq.~\eqref{eq:delta}. We set $\Omega T = 1$ and the detector is at (dimensionless) proper distance $d=1$ from the horizon. (a) Small mass regime $M/T=1$. (b) Large mass regime $M/T = 100$. In all cases the deformation increases the detector responses, but they grow with different scaling behaviour. }
    \label{fig: Fig2-coupling}
\end{figure}

In Figure~\ref{fig: Fig2-coupling} we compare how the deformation parameters $l,\alpha,a$ (which after rescaling with $T$, we collectively denoted by $\delta$)
of the different metrics change the detector response relative to Schwarzschild. Here we define the dimensionless parameter
\begin{align}
    \label{eq:delta}
    \delta = \begin{cases}
    l/T \quad &\text{(Hayward)}\\
    \alpha/T^2  &\text{(EGB)}\\
    a/T &\text{(Black bounce)}
    \end{cases}
\end{align}
for convenience. We see that the detector response increases with larger deformation in all cases although they grow at different rates: the EGB case in particular grows very quickly, followed by the BB, the slowest being Hayward. For the small mass regime, the Hayward deformation $l$ has the largest effect, while for BB and EGB their impact only becomes apparent when the parameters $\alpha,a$ become rather large. Note that for $M/T=1$ in Figure~\ref{fig: Fig2-coupling}(a), we have restricted $\delta\leq 0.76$ since the Hayward metric limits the size of $l$ according to the critical mass $m_c$ in Eq.~\eqref{eq: critical-hayward};  similarly for EGB (there is no upper bound for $a$ in BB case). In Figure~\ref{fig: Fig2-coupling}(b), however, the large mass $M/T=100$ allows $\delta$ to take large values to make sense, so $\delta\leq 10$ is in fact ``small deformation''. In any case, Figure~\ref{fig: Fig2-coupling} shows that we can indeed understand the potential impact of higher curvature gravity and non-standard static spherically symmetric metrics and obtain qualitative picture of how corrections to the detector response would look like.

\begin{figure}[tp]
    \centering
    \includegraphics[scale=0.74]{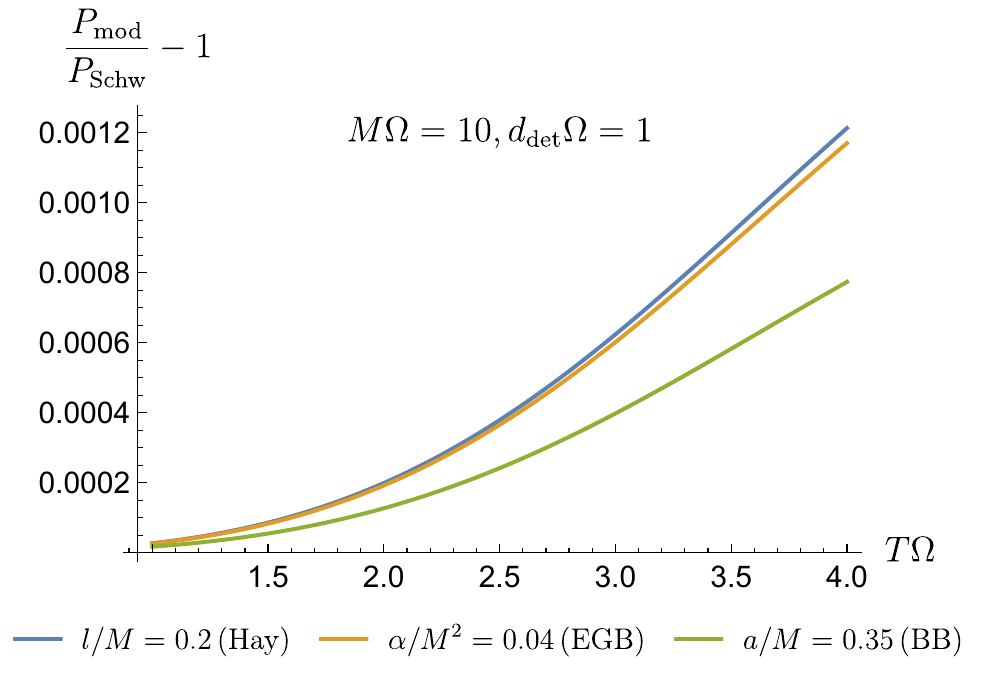}
    \includegraphics[scale=0.74]{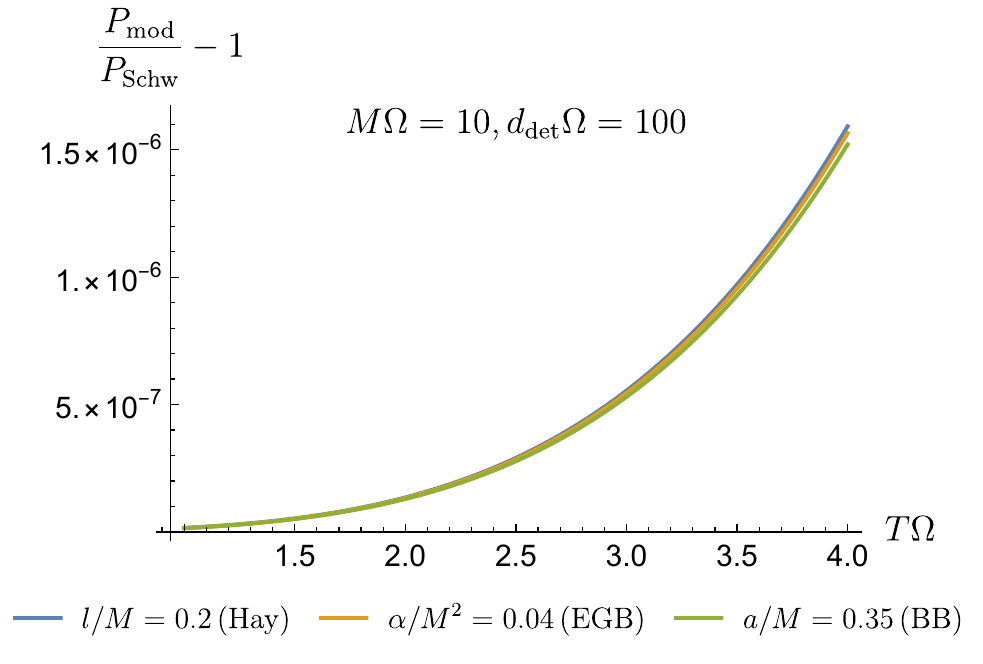}
    \caption{The probability differences, computed as ratio $\mathcal{R}=P_{\text{mod}}/P_{\text{Schw}}-1$ as a function of switching time $T$. All quantities are measured in units of $\Omega$, but we keep the deformation parameter to be scaled with $M$ (which can be converted to $\Omega$ if one likes). We set the mass to be $M\Omega = 10$. (a) Near horizon regime, with proper distance $d_\text{det}\Omega=1$. (b) Far from black hole regime, with proper distance $d_\text{det}\Omega=100$. The detector response is more sensitive near the black hole in distinguishing the metric functions. In all cases, longer interaction improves the sensitivity of the detector to the deformation parameter.}
    \label{fig: Fig3-time}
\end{figure}

\begin{figure}[tp]
    \centering
    \includegraphics[scale=0.75]{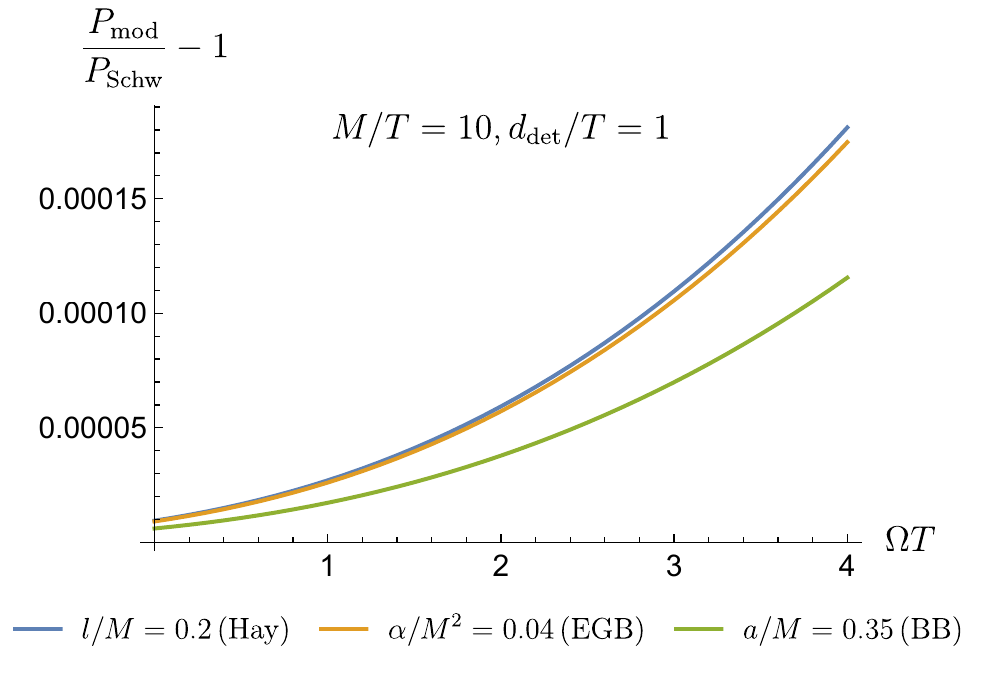}
    \includegraphics[scale=0.75]{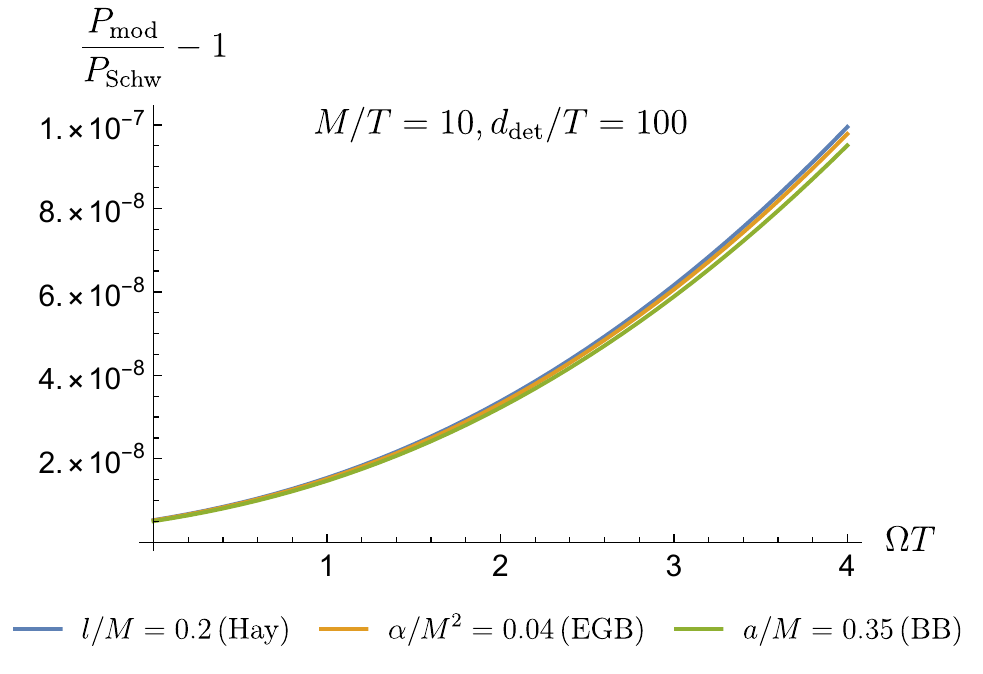}
    \caption{The probability differences, computed as ratio $\mathcal{R}=P_{\text{mod}}/P_{\text{Schw}}-1$ as a function of energy gap $\Omega$. This time, all quantities are measured in units of $T$, but we keep the deformation parameter to be scaled with $M$. We set the mass to be $M/T = 10$. (a) Near horizon regime, with proper distance $d_\text{det}/T=1$. (b) Far from black hole regime, with proper distance $d_\text{det}/T = 100$. As before, the detector response is more sensitive near the black hole in distinguishing the metric functions. In all cases, larger energy gap improves the sensitivity of the detector to the deformation parameter. Although qualitatively similar to Figure~\ref{fig: Fig3-time}, the scales are different.}
    \label{fig: Fig4-gap}
\end{figure}

We close this section by analysing how the detector response behaves as the energy gap and interaction time is varied. The basic idea is that with longer interaction time, the detector can sample more curvature variations and we expect that for any fixed deformation away from Schwarzschild, longer interaction enables the detector to distinguish the metric functions better. For the  Hayward and EGB cases in particular, while we know that $\alpha=l^2$ makes their small-$l$ (resp. small-$\alpha$) expansion identical to leading order, longer interaction will effectively probe sub-leading term in the expansion, thus distinguishing the two metrics. Indeed this is the case as we show in Figure~\ref{fig: Fig3-time}.  Similar behaviour can be expected from varying energy gap: larger energy gap probes the short-distance part of the field, which can then resolves the differences between the metric functions better; this is shown in Figure~\ref{fig: Fig4-gap}. Thus we expect that Fig~\ref{fig: Fig3-time} and \ref{fig: Fig4-gap} are qualitatively similar.  Note that for both variations in interaction time and energy gap, being further away from the black hole makes detector response differences smaller: the expansion in the parameters $l,\alpha,a$ in Eq.~\eqref{eq: small-expansion} decay as $O(r^{-3})$, which is much faster than the ``Schwarzschild part'' of the metric function, so they contribute less to the detector response at large distances.

\section{Thermality in a  static spherically symmetric background}
\label{sec: thermality}

In this section we will show that our generalization is good enough to capture the standard formula for the  (Gibbons-)Hawking temperature obtained in the literature on thermodynamics of black holes and cosmological horizons. That is, we can define a notion of thermality and detailed balance condition using the derivative coupling Wightman two-point function for arbitrary static spherically symmetric geometries with Killing horizons.

\subsection{KMS condition}
\label{sec: pseudo-KMS}

The \textit{Kubo-Martin-Schwinger} (KMS) \textit{condition} gives a formal statement about thermalization in quantum field theory  \cite{Kubo1957thermality,Martin-Schwinger1959thermality,Haag1967KMS}. For scalar field theory, the condition implies that the vacuum state of the field is thermal with respect to the timelike vector field $\partial_\tau$ if the vacuum Wightman two-point distribution satisfies complex anti-periodicity
\begin{align}
    W(\tau,\tau'+\ii\beta) = W(0,-(\tau-\tau'))\,.
\end{align}
Since the Wightman function is stationary with respect to $\tau$, this reduces to $W(\Delta\tau,\ii\beta) = W(0,-\Delta\tau)$ where $\Delta\tau = \tau-\tau'$. The KMS temperature is given by $\mathsf{T}_{\textsc{KMS}} = \beta^{-1}$. 

In the derivative coupling model, the notion of a KMS condition  only makes sense phenomenologically, since the proper-time derivative of the scalar field operator-valued distribution $\partial_\tau\hat\phi(\sx(\tau))$ is not really an element of the algebra of observables of the theory\footnote{
One can potentially include proper-time derivatives into the algebra the same way one can enlarge the algebra by including the stress-energy tensor, but for our purposes it is not necessary to take this step, and we may content ourselves with calling this pseudo-KMS condition if we prefer.}. However, the derivative coupling UDW model does exhibit a Planckian thermal response when it undergoes uniform acceleration in flat space or is on a static trajectory in black hole spacetimes \cite{Aubry2014derivative,Aubry2018Vaidya,tjoa2020harvesting}. Therefore, one may hope to be able to construct some sort of ``pseudo-KMS'' property just for the derivative coupling model so that the Planckian response can be interpreted as a consequence of the pseudo-KMS property\footnote{Alternatively, as mentioned earlier one can view this as enlarging the algebra of observables to include proper-time derivatives of the field.}. We will try to see how this can be formulated and under what circumstances it will make sense.  For convenience we simply refer this as the KMS condition.

First, let us forget about black holes and check what happens for an accelerating detector in \textit{flat} spacetime. The standard Wightman function is given by
\begin{align}
    W_\text{flat}(\sx,\sx') = -\frac{1}{4\pi}\log\left[-\Lambda^2(\Delta u - \ii\epsilon)(\Delta v-\ii\epsilon)\right]\,,
\end{align}
where $\Delta u = u-u',\Delta v = v-v', u = t-x,v = t+x$. For an accelerating detector, we have
\begin{align}
    t(\tau ) = \frac{1}{a}\sinh(a\tau)\,,\hspace{0.5cm}x(\tau) = \frac{1}{a}\cosh(a\tau)\,.
\end{align}
Together, these give us a derivative coupling Wightman distribution:
\begin{align}
    \mathcal{A}_\text{flat}(\tau,\tau') = -\frac{1}{4\pi}\left[\frac{\dot{u}\dot{u}'}{(u-u'-\ii\epsilon)^2} +\frac{\dot{v}\dot{v}'}{(v-v'-\ii\epsilon)^2}  \right]\,.
\end{align}
This is formally the same exact form as  for the  Boulware vacuum \eqref{eq: derivative-boulware}. Note that the numerator potentially breaks stationarity since for an accelerating detector $\dot{u}$ (resp. $\dot{v}$) is a function of $\tau$. This case is important because the same will occur for the Hartle-Hawking vacuum: we have $\dot{U} \neq \text{constant}$.

It turns out that for an accelerating detector with constant proper acceleration $a$, the properties of hyperbolic sine and cosine conspire to produce a \textit{stationary} Wightman function:
\begin{align}
    \mathcal{A}_{\text{accel}}(\tau,\tau') = -\frac{a^2}{8\pi}\text{csch}^2\left[\frac{a(\tau-\tau')}{2}\right]\,,
    \label{eq: derivative-accel}
\end{align}
where $\text{csch}(z) = 1/\sinh(z)$ is the hyperbolic cosecant. We removed the $\ii\epsilon$ for clarity and interpret it as an instruction to perform the integral in $P(\Omega)$ along certain contours. The key observation to be made here is that up to a fixed constant prefactor (off by a factor of $1/(2\pi)$), this is \textit{exactly} the same as the expression for the Unruh effect in (3+1)-dimensional flat spacetime. Therefore, in this case the apparent lack of stationarity in the form of $\mathcal{A}_\text{flat}$ does not prevent the derivative coupling Wightman function from being stationary. This gives us hope that indeed we can formulate  the KMS condition proper
for both the Unruh and Hawking effects in the derivative coupling model.

Let us now try it for the generalized Hartle-Hawking vacuum for SSS spacetimes with a  single metric function $f(r)$. For a stationary trajectory, we have $r_\star = r_\star(r) = \text{constant}$ since $r$ is fixed, and $t(\tau) = \tau/\sqrt{f(r)}$. Consequently, we have
\begin{subequations}
\begin{align}
    u(\tau) &= \frac{\tau}{\sqrt{f(r)}} - r_\star(r)\,,\\
    v(\tau) &= \frac{\tau}{\sqrt{f(r)}} + r_\star(r)\,,\\
    \dot{U}(\tau) &= e^{-\kappa_\textsc{h}u(\tau)}\dot{u}(\tau) =  \frac{1}{\sqrt{f(r)}}e^{-\kappa_\textsc{h} u(\tau)}\,,\\
    \dot{V}(\tau) &= e^{\kappa_\textsc{h}v(\tau)}\dot{v}(\tau) =  \frac{1}{\sqrt{f(r)}}e^{\kappa_\textsc{h} u(\tau)}\,.
\end{align}
\end{subequations}
Substituting the generalized Kruskal coordinates, the derivative coupling Wightman distribution reads
\begin{align}
    \mathcal{A}_\textsc{h}(\tau,\tau') &= -\frac{1}{4\pi}\frac{1}{f(r)}\left[\frac{e^{-\kappa_\textsc{h}u(\tau)}e^{-\kappa_\textsc{h}u(\tau')}}{\left[e^{-\kappa_\textsc{h}u(\tau)}-e^{-\kappa_\textsc{h}u(\tau')}\right]^2} + \frac{e^{\kappa_\textsc{h}v(\tau)}e^{\kappa_\textsc{h}v(\tau')}}{\left[e^{\kappa_\textsc{h}v(\tau)}-e^{\kappa_\textsc{h}v(\tau')}\right]^2}\right] \notag\\
    &= -\frac{\kappa_\textsc{h}^2}{8\pi f(r)}\text{csch}^2\left[\frac{\kappa_\textsc{h}(\tau-\tau')}{2\sqrt{f(r)}}\right]\,.
    \label{eq: derivative-radiation}
\end{align}
Remarkably, this is exactly the same as the derivative coupling variant of the Unruh effect \eqref{eq: derivative-accel}, with the identification $a\to \kappa_\textsc{h}/\sqrt{f(r)}$.\footnote{The extra factor $f(r)$ enforces the condition that the detector cannot remain static when $f(r)\to 0$.} Therefore, the derivative coupling Wightman function $\mathcal{A}_\textsc{h}(\tau,\tau')$ is both stationary in $\tau,\tau'$ and obeys the KMS condition:
\begin{align}
    \mathcal{A}_\textsc{h}(\Delta\tau,0) = \mathcal{A}_\textsc{h}(\tau,\tau')\,,\quad \mathcal{A}_\textsc{h}(\Delta\tau+\ii\beta,0) = \mathcal{A}_\textsc{h}(0,-\Delta\tau)\,.
    \label{eq: pseudo-KMS}
\end{align}
The KMS temperature $T_\textsc{h}$ is obtained from the fact that the imaginary (anti-)periodicity is given by $\ii\beta\kappa_\textsc{h}/(2\sqrt{f(r)}) = 2\pi \ii$. That is, the KMS temperature is given by the well-known formula for black hole horizons (or cosmological horizons) \cite{Gibbons1977cosmological,hawking1975particle,Kubiznak2017chemistry}, but corrected by a gravitational redshift factor (also known as \textit{local Tolman temperature} $\mathsf T_\textsc{loc}$) \cite{Tolman1930weight-heat,TolmanEhrenfest1930temperature}:
\begin{align}
    \mathsf{T}_\textsc{loc} \coloneqq \frac{1}{\sqrt{f(r)}} \mathsf{T}_\textsc{h}\,,\quad \mathsf T_\textsc{h} = \beta^{-1} = \frac{\kappa_\textsc{h}}{2\pi} = \frac{f'(r_\textsc{h})}{4\pi}\,.
    \label{eq: Hawking-temperature}
\end{align}
We have thus shown that two-dimensional truncation of the derivative coupling two-point function for \textit{arbitrary} static spherically symmetric metric with metric function $f(r)$ is a thermal KMS two-point function with respect to $\partial_\tau$ with temperature given by the \textit{local} Hawking temperature.

\subsection{Detailed balance condition}
\label{subsec: detailed-balance}

Finally, we can now   straightforwardly formulate the \textit{detailed balance condition} for two-dimensional truncation of arbitrary truncation of SSS metric. The idea is to use the fact that in the case of Unruh effect in arbitrary $(3+1)$-dimensional Minkowski space, the detailed balance condition is given by\footnote{This is typically written by rescaling the coupling strength with the timescale of interaction and take the long time limit. For Gaussian switching, this procedure is the same as simply taking the ratio of the probabilities directly and compute the late-time limit.} \cite{Fewster2016wait,Pipo2019without}
\begin{align}
    \lim_{T\to\infty}\frac{P(\Omega)}{P(-\Omega)} = \frac{\widetilde{W}_\text{accel}(\Omega)}{\widetilde{W}_\text{accel}(-\Omega)} = e^{-\beta_{\text{accel}}\Omega}\,.
    \label{eq: detailed-balance-accel}
\end{align}
Here $\widetilde{W}_\text{accel}(\Omega)$ is the Fourier transform of the Wightman two-point function ${W}_\text{accel}(\tau,\tau')$ for accelerating trajectory in (3+1)-dimensional flat space. However, since $\mathcal{A}_\text{accel}$ in Eq.~\eqref{eq: derivative-accel} is identical to ${W}_\text{accel}$ up to a constant prefactor, the ratio of the Fourier transform is identical, so derivative coupling variant will also obey the \textit{exact} same detailed balance condition \eqref{eq: detailed-balance-accel}. This agrees with the transition rate calculation obtained in \cite{Aubry2014derivative}.

For static spherically symmetric geometries with horizons,  we saw earlier from \eqref{eq: derivative-radiation} that $\mathcal{A}_\textsc{h}$ is identical to $\mathcal{A}_\text{accel}$ but with the replacement $a\to \kappa_\textsc{h}/\sqrt{f(r)}$. Since the trajectory is \textit{stationary} as the detector is at constant $r$, $\kappa^2_\textsc{h}/f(r)$ is constant and hence the Fourier transform of $\mathcal{A}_\textsc{h}(\tau,\tau')$ is  \textit{identical} to the case of Unruh effect. The detailed balance condition is therefore identical to \eqref{eq: detailed-balance-accel}, but with a minor replacement
\begin{align}
    \lim_{T\to\infty}\frac{P(\Omega)}{P(-\Omega)} = \frac{\widetilde{\mathcal{A}}_\textsc{h}(\Omega)}{\widetilde{\mathcal{A}}_\textsc{h}(-\Omega)} = e^{-\beta_{\textsc{loc}}\Omega}\,.
    \label{eq: detailed-balance-blackhole}
\end{align}
The difference now is that since the identification is $a\to \kappa_\textsc{h}/\sqrt{f(r)}$ and not just $\kappa_\textsc{h}$, the RHS is not the  $\beta_\textsc{h}$ associated with the Hawking temperature $\mathsf{T}_\textsc{h}$ but with the {local} Tolman temperature $\mathsf{T}_\textsc{loc}$ \cite{Tolman1930weight-heat,TolmanEhrenfest1930temperature}. 
This reduces to the standard result found in \cite{birrell1984quantum} and numerical analysis in \cite{tjoa2020harvesting} when we specialize to the Schwarzschild metric with  $f(r) = 1-2M/r$.

\section{Discussion, outlook and open challenge}

We have shown that, provided one does not ask questions that require angular directions in important ways (such as what happens to UDW detectors in stationary orbits), the derivative coupling variant of the UDW model admits a much larger generalization to include arbitrary static spherically symmetric geometries from \textit{any theory of gravity}, or metrics that do not come from standard solutions to the Einstein equations, such as regular black hole solutions of the Hayward and the black bounce metrics. We were able to show, for instance, that detector response are sensitive to the $O(r^{-3})$ fall-off behaviour of the gravitational field (which is not present in standard RN-(A)dS solutions, cf. Eq.~\eqref{eq: RN-AdS}), and this sensitivity improves with longer interaction, larger detector gap, and being closer to the black hole. Our work provides a modest and accessible way to understand what the Unruh-DeWitt model and standard methods in QFT in curved spacetimes can say about the impact of modified gravity theory and metric deformations away from the standard Schwarzschild solution.

We have also shown that this generalization still respects the KMS property, in that the corresponding generalized Hartle-Hawking state yields a Wightman function that is thermal with respect to the KMS temperature given by a well-known formula in black hole thermodynamics, namely $\mathsf T_\textsc{h} = f'(r_\textsc{h})/(4\pi)$. This formula is typically obtained using the ``Euclidean'' trick and near-horizon expansion, or from path integral considerations. We have shown that the 2D truncation allows us to obtain the same temperature at the level of the (derivative coupling) Wightman function, generalizing previously known results \cite{Aubry2014derivative,tjoa2020harvesting}). 

We remark that while we have picked a particular modified gravity theory and some non-standard solutions in general relativity, our construction should be applicable to a wide class of modified gravity theories that exist in (3+1) dimensions, such as scalar-tensor and Horndeski theories and their static spherically symmetric solutions. Similarly, there exists non-standard solutions due to non-standard matter content, such as dyonic black holes (possibly due to non-linear electrodynamics), metrics that are motivated from certain limits of string theory \cite{dijgraaf1992string,witten1991stringBH}, or black holes from Einstein-Yang-Mills theory with non-Abelian charges \cite{Kleihaus2002EYM}. We have also not investigated what happens to \textit{infalling} observers, since some of the non-standard metrics (especially regular black hole solutions) are very different from Schwarzschild in the interior geometry as compared to their exterior geometry. As this is numerically challenging and the goal in this paper is more on the formalism, we leave this for future investigations.

We hope that our results will enable more effort in trying to understand QFT in curved spacetimes when the metric describing the curved backgrounds has  non-standard origins. Before   closing, below we discuss some open questions and what we think may be worth further investigation. For concreteness, we pose some open challenges for the community who may be interested in some of these extensions.

\subsection{Higher derivative coupling?}

One natural question that comes to mind is whether the derivative coupling itself admits higher-derivative generalizations. This is because the derivative coupling model essentially replicates the short-distance behaviour of Hadamard states in (3+1) dimensions in appropriate regimes but \textit{not} any higher dimensions. It is actually not difficult to check that na\"ive generalization by taking more proper-time derivatives will \textit{not} work. As a simple example, even for accelerating detectors in flat space, the double-proper-time derivative would have been associated with a 2D truncation of the Wightman function in $(5+1)$ dimensions; however  direct computation shows that
\begin{align}
    \mathcal{A}^{(2)}_{\text{accel}}(\tau,\tau') = \partial_{\tau}^2\partial_{\tau'}^2W(\sx(\tau),\sx(\tau')) \propto \frac{\cosh (a(\tau-\tau'))+2}{\sinh^4\left(\frac{a(\tau-\tau')}{2}\right)}\,.
\end{align}
This expression still has the required thermal properties as it respects the (anti-)periodicity in imaginary time because the extra $\cosh(a\Delta\tau)$ is an even function and the holomorphicity property in the $\ii\beta$-strip is not really spoiled by the numerator\footnote{See \cite{Pipo2019without} for more discussion on properties of the KMS condition and in particular the holomorphicity of the Wightman function.}. However, this does not have the correct Hadamard short-distance property because $\cosh(a\Delta\tau)$ in the numerator effectively ``screens'' the large-$\Delta\tau$ regime (weakening the fall-off at large $\Delta \tau$) while enhancing the singular part in the  $\tau\approx \tau'$ regime by an approximately constant shift (since $\cosh a\Delta\tau\approx 1$). Therefore, \textit{at best} this generalization would only work when $\Delta\tau$ is small (rescale the constant shift), but for physics of thermalization we need $\Delta\tau$ to be large.

Note also that even if the higher-derivative coupling \textit{were} to work (and we just argued above that it did not), it would have covered only higher even-dimensional spacetimes but not odd-dimensional ones. One can therefore ask how to construct derivative-coupling variant that replicates the properties of Wightman functions in (2+1)-dimensional spacetimes. Two of the most well-studied classes of non-flat background in this dimension are AdS$_3$ and the Ba\~nados-Teitelboim-Zanelli (BTZ) black holes \cite{BTZ1993,Lifschytz1994BTZ}. While the QFT in (2+1) dimensions is more tractable and in some cases (like BTZ black holes)  can be analytically solved to a large extent\footnote{The issue here is that unlike in (1+1) dimensions, the vanishing of the Weyl tensor vanishes in (2+1) dimensions still does not guarantee conformal flatness: one needs the Cotton tensor to vanish.}, { higher dimensional generalizations of the BTZ spacetime have been studied to some extent \cite{deSouzaCampos:2020bnj}, and
we may wonder if we can construct a (1+1) derivative-type model that can mimic properties of the (2+1) UDW model and its higher odd-dimensional generalizations.   }

One natural guess would be to consider ``fractional derivatives'' (see, e.g., \cite{Khalil2014fractional} for background and the framework of fractional calculus). For example, if we have a ``half-derivative'' $D^{1/2}$ of a function $f$, applying this twice will recover standard first derivative of $f$. However, it can be checked that one particular definition, the  \textit{Riemann-Liouville fractional derivative}, defined by (for one-variable case) \cite{Khalil2014fractional}
\begin{align}
    D^{\alpha}_bf(x)\coloneqq \frac{1}{\Gamma(1-\alpha)}\frac{\dd}{\dd t}\int_b^x\dd x\frac{f(t)}{(x-t)^\alpha}\,,\quad \alpha\in [0,1)\,,
\end{align}
will ``almost'' reproduce (2+1)-dimensional short-distance behaviour but it suffers similar screening/enhancement effects as the double-time derivative considered earlier.

While we do not exhaust all possible fractional derivative definitions, we leave this interesting generalization for the future. We open the challenge for finding a version of fractional differentiation that works, or else to prove that such methods will never work because of the unavoidable screening/enhancement effects.

\subsection{Maximal extension of derivative coupling model?}

Assuming that derivative coupling variant only works for (3+1)-dimensional truncation, our work here shows that in some sense generalization to an arbitrary metric function $f(r)$ is ``maximal'', i.e., there is no further (natural) extension that the derivative coupling variant can cover. This was our original motivation, which is to exhaust the utility of the model with respect to the (truncated) curved background on which it could be applied. We emphasize that if the metric function $f(r)$ comes from some arbitrary theory of gravity, then the theory has to at least have  non-trivial dynamics in (3+1) dimensions that differ from general relativity. This is the reason why we needed 4D EGB gravity instead of the more familiar and extensive class of Lovelock gravity, whose higher-curvature corrections only matter in five dimensions or higher.

Note that the derivative coupling variant is in a sense a ``sleight of hand'':   a clever choice of coupling in the  (1+1)-dimensional calculation captures some important features of (3+1)-dimensional physics (such as the Unruh and Hawking effects). Since there are no Einstein equations in (1+1)-dimensions, the two-dimensional UDW model and QFT in curved spacetimes do not generally have a semi-classical regime where matter can backreact to the metric (unless one considers
a two-dimensional limit
of general relativity
\cite{Mann:1992ar,Sikkema:1989ib} or some 
modified theory such as Jackiw-Teitelboim gravity \cite{Mann1990JT}; see \cite{Blommaert2021JT-UDW} for a recent UDW application in this context). Therefore, one advantage of using the  (1+1)-dimensional derivative UDW model is that whether or not the source of the truncated metric comes is from some gravitational field equations in two dimensions is irrelevant.

Since the truncated metric is non-dynamical (and is not required to be a solution of some two-dimensional theory), one can ask if there is a clever, even if \textit{ad hoc}, trick to incorporate the effect of rotation (angular momentum) such as that of Kerr black hole\footnote{Note that one needs to care about ``which'' two-dimensional slice of a rotating black hole one chooses to work with, since the causal structure of a rotating black hole also depends on the angle. We thank Robie A. Hennigar for pointing this out.}, or the effect of gravitational waves. Since there are not enough spatial dimensions, there can be no rotating black hole in (1+1) dimensions nor gravitational waves. However, it was recently pointed out \cite{Gray2021imprint} that there are still imprints of (non-linear) gravitational shockwaves on the Wightman function even when one specializes to the case of a pointlike  detector   without any extension in the transverse dimension. Since  a gravitational shockwaves \textit{is} a non-linear gravitational wave, perhaps it is possible to incorporate the effect of gravitational waves in (1+1) dimensions in some manner, with the dependence on the transverse dimension treated as an ``external parameter'' of the Wightman function. 

A future challenge is to construct  a \textit{systematic} (not case-by-case, although this itself is interesting) procedure that allows one to encode the effects of rotation and gravitational waves as a external parameters of the two-dimensional Wightman function: if this is possible in general, then it would extend our construction to these situations. As an explicit example, if rotations were possible to encode, one might  even study rotating or NUT-charged black holes in higher curvature gravity, including more exotic ones such as the Gauss-Bonnet-Taub-NUT solution in \cite{hennigar2020taking}.

\section*{Acknowledgment}
This work was supported in part by the Natural Sciences and Engineering Research Council of Canada.
The authors thank Finnian Gray for useful discussions and Robie A. Hennigar for reading the draft of the manuscript and providing useful feedback for improvement. E. T. acknowledges generous support from the Mike and Ophelia Lazaridis Fellowship, as well as the (especially mental/psychological) support of his supervisors Robert B. Mann and Eduardo Mart\'in-Mart\'inez during this work. This work is conducted on the traditional territory of the Neutral, Anishnaabeg, and Haudenosaunee Peoples. The University of Waterloo and the Institute for Quantum Computing are situated on the Haldimand Tract, land that was promised to Six Nations, which includes six miles on each side of the Grand River.

\bibliography{Vaidyabib}
\bibliographystyle{JHEP}

\appendix

\end{document}